\documentclass[a4paper]{spie}  
\usepackage[]{graphicx}
\usepackage{cite} 
\usepackage{amsmath}

\title{Predicting the whispering gallery mode spectra of microresonators} 

\author{Jonathan M. M. Hall,\supit{a} Shahraam Afshar V.,\supit{a} Matthew R. Henderson,\supit{a} Alexandre Fran\c{c}ois,\supit{a} Tess Reynolds,\supit{a} Nicolas Riesen\supit{a} and Tanya M. Monro\supit{a,b}
\skiplinehalf
\supit{a}Institute for Photonics \& Advanced Sensing and ARC Centre for Nanoscale BioPhotonics, School of Physical Sciences, The University of Adelaide, Adelaide, SA 5005, Australia; 
\skiplinehalf
\supit{b}The University of South Australia, Adelaide, SA 5000, Australia
}

\authorinfo{Further author information: (Send correspondence to J.M.M.H.)\\J.M.M.H.: E-mail: jonathan.hall@adelaide.edu.au}

\begin{document} 
  \maketitle 

%%%%%%%%%%%%%%%%%%%%%%%%%%%%%%%%%%%%%%%%%%%%%%%%%%%%%%%%%%%%% 
\begin{abstract}
The whispering gallery modes (WGMs) of optical resonators have prompted intensive research efforts due to their usefulness in the field of biological sensing, and their employment in nonlinear optics. While much information is available in the literature on numerical modeling of WGMs in microspheres, it remains a challenging task to be able to predict the emitted spectra of spherical microresonators.
Here, we establish a customizable Finite-Difference Time-Domain (FDTD)-based approach to investigate the electromagnetic WGM spectrum of microspheres. The simulations are carried out in the vicinity of a dipole source rather than a typical plane-wave beam excitation, thus providing an effective analogue of the fluorescent dye or nanoparticle coatings used in experiment.
The analysis of a single dipole source at different positions on the surface or inside a microsphere, serves to assess the relative efficiency of nearby radiating TE and TM modes, characterizing the profile of the spectrum.
By varying the number, positions and alignments of the dipole sources, different excitation scenarios can be compared to analytic models, and to experimental results.
The energy flux is collected via a nearby disk-shaped region. 
The resultant spectral profile shows a dependence on the configuration of the dipole sources.
The power outcoupling can then be optimized for specific modes and wavelength regions. The development of such a computational tool can aid the preparation of optical sensors prior to fabrication, by preselecting desired the optical properties of the resonator. 
\end{abstract}

\keywords{Whispering gallery modes, Microresonators, Resonators, Optical devices, Microcavities, Computational optics, Optics modeling, Spectroscopy}

%%%%%%%%%%%%%%%%%%%%%%%%%%%%%%%%%%%%%%%%%%%%%%%%%%%%%%%%%%%%%
\section{INTRODUCTION}
\label{sec:intro} 

Whispering gallery modes (WGMs) are produced by the propagation of electromagnetic waves along  
the interface of the surface of a resonator and its environment. As a result, 
an evanescent field is produced, which extends beyond the 
surface of the resonator. 
WGMs represent an optical phenomenon which is applicable to sensing applications, due to the 
sensitivity of the evanescent field to the presence  
of virions \cite{Vollmer2008591}, and 
macromolecules such as proteins 
\cite{Vollmer:2002a,Boyd:01,Arnold:03,Ksendzov:05}, which 
break the symmetry of the field \cite{doi:10.1021/nl401633y}.  
WGMs have been used 
for the development of label-free detection technologies \cite{Armani:2007a,Baaske2014}, 
optical frequency combs \cite{Liang:11}, 
nonlinear optics \cite{Schliesser2010207} and quantum electrodynamics (QED) 
\cite{PhysRevA.83.063847}. 
This work primarily focuses on the establishment of a computational 
modeling tool for investigating properties of WGM spectra in microspheres and microshells. 

Optical microresonators have been the subject of much recent investigation 
since the discovery of extremely high quality($Q$)-factor generation ($> 10^8$), 
which is possible by exciting the WGMs of a resonator, using a prism, 
or a tapered fiber coupled to the resonator 
\cite{Knight:97,Hossein-Zadeh:06,1674-1056-17-3-047,s100301765,6525394}. 
An alternative approach to this is to excite the WGMs indirectly, using fluorescent 
media \cite{Francois:13}, surface plasmon effects  
\cite{Armani:2007a,Min:2009a} or nanoparticle coating/doping using  
quantum dots \cite{Shopova:2003a}. 
WGMs of the resonator also exhibit high sensitivity to their external environment. 
This means that not only are WGMs potentially very distinguishable
 from one another in the power spectrum, but that changes due to the presence of a new electromagnetic source can also be identified easily. 

The manufacture of optical resonators suited to a particular purpose using trial-and-error methods is expensive. Efforts to simulate and characterize the optical properties of WGMs in resonators using computational methods have so far focused on specific and ideal scenarios, such as excitation from a plane-wave beam \cite{Fujii:2005a,Fujii:2005b,Quan:2005a}. This opens the way for an investigation into realistic mode excitation scenarios, and analyzing differences in the spectral profiles, with a view to
making a robust prediction tool. 
Here, we investigate coupling strategies that involve 
the excitation of WGMs of microresonators,  
using dipole sources placed on the surface. 
The Finite-Difference Time-Domain (FDTD) 
computational tool developed in this work will address 
this coupling scenario in particular. 
The tool is also easily extended to include 
fiber-coupled methods of mode excitation.

\section{FDTD MODEL}

%RP
The FDTD technique \cite{taflove1995computational} splits a volume of
space into a three-dimensional grid, and calculates the solution of Maxwell's Equations
for the electromagnetic fields at each point. This is repeated for a finite number of
time increments, and performed sequentially as the system evolves in time. FDTD
constitutes a comprehensive solution of an optical system, with all radiation modes
present in the simulation. The only artifacts in the calculation 
involve discretization and finite-volume effects, and assumptions made regarding 
ideal material properties and resonator configurations. 
The FDTD method is also suitable for accessing transient or unexpected electromagnetic
properties, as each time step is evaluated individually. However, the method is
computationally expensive when achieving a result that is sufficiently
converged~\cite{2014arXiv1410.8608H}. 

The power spectrum represents a useful quantity for representing the positions and the $Q$-factors
of the excited modes \cite{Henderson:11}. To obtain this, a flux-collection region may be introduced into the 
simulation, placed a certain distance from the microresonator. 
The total output power ($P$) in terms of wavelength 
($\lambda$) is obtained by integrating the projection of the Poynting vector 
($\mathbf{S} \equiv \mathbf{E} \times \mathbf{H}^*$) onto a  
flux region of area $A$, with a normal unit vector $\hat{\mathbf{n}}$ 
\begin{equation}
P(\lambda) = \int\!\mathbf{S}\cdot \hat{\mathbf{n}}\,\, \mathrm{d}A. 
\end{equation}
The profile of the power spectrum, plotted as a function of $\lambda$, 
will show sharp 
peaks that correspond to the positions of TE and TM WGMs. 
 Changes in this energy output exactly map out the WGMs,
and thus offer an excellent diagnostic tool for the quality of the resonator scenario
being examined.

An important aspect in developing the FDTD tool is a comparison with typical analytic models 
used in the field \cite{Johnson:93,Teraoka:06a}, 
which can provide estimates on the systematic uncertainty of the WGM positions. 
To check the mode positions predicted by FDTD, an analytic model is used, in
which Maxwell's Equations are solved in a dielectric medium with a boundary \cite{Johnson:93}. 
In this model, the resonance conditions for the TE and TM modes may be expressed in terms of the spherical 
Bessel functions 
of the first kind, $j_m$, and the spherical Hankel function, $h_m^{(1)}$, for azimuthal mode number $m$. 
These conditions are 
\begin{align}
 \frac{n_1}{z^R_1} \frac{(m+1) j_m(z^R_1) - z^R_1\, j_{m+1}(z^R_1)}{j_m(z^R_1)} &= 
\frac{n_2}{z^R_2} \frac{(m+1) h_m^{(1)}(z^R_2) -  z^R_2\, h^{(1)}_{m+1}(z^R_2)}{h^{(1)}_m(z^R_2)}. \\
\mbox{and}\quad \frac{1}{n_1 z^R_1} \frac{(m+1) j_m(z^R_1) - z_1\, j_{m+1}(z^R_1)}{j_m(z^R_1)} &= 
\frac{1}{n_2 z^R_2} \frac{(m+1) h^{(1)}_m(z^R_2) -  z_2\, h^{(1)}_{m+1}(z^R_2)}{h^{(1)}_m(z^R_2)},
\end{align}
respectively, for a refractive index of $n_1$ inside a sphere of radius $R$, and $n_2$ outside the sphere.  
The size parameters are defined as $z^R_{1,2} \equiv   k\, R\, n_{1,2}$, for a wave number $k=\omega/c$, 
where $c$ is the speed of light in a vacuum.  
Accurate matching of the mode structure of the spectrum allows one to identify the dominant contributions to the features of the spectrum. This identification facilitates future experimental studies in the reduction or enhancement of these features.

\begin{figure}
\centering
\includegraphics[height=160pt]{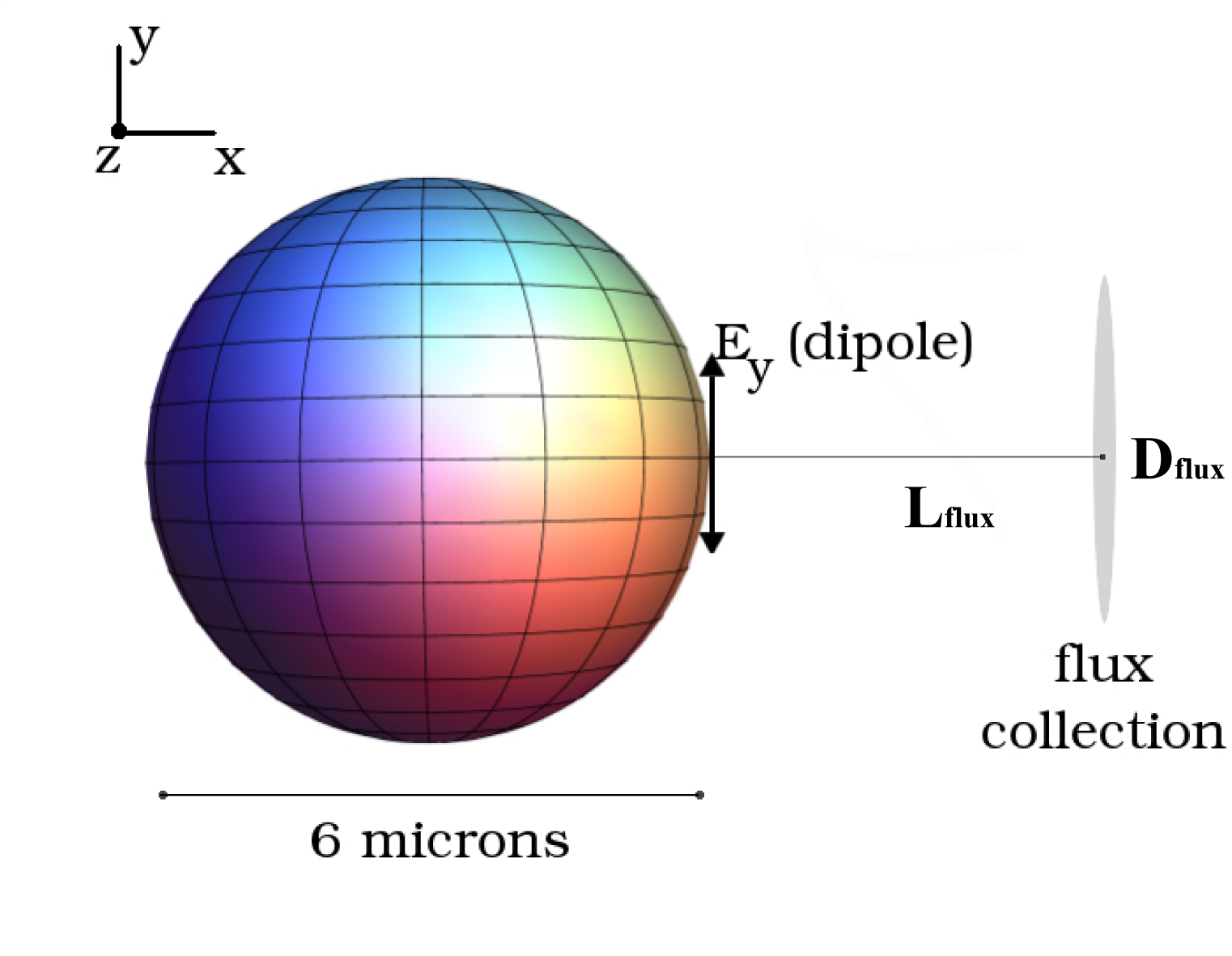}
\vspace{-3mm}
\caption{\footnotesize{ A circular flux collection region, offset from a $6$ $\mu$m diameter microsphere, is placed so that its normal is aligned perpendicular to the surface of the sphere. In this example illustration, the optical modes are excited from a tangentially oriented electric dipole source. 
}}
\label{fig:Sphere}
\end{figure}

\section{SIMULATION RESULTS}

The FDTD method 
will be used to consider a variety of flux collection scenarios, 
source distribution configurations, and resonator properties. 
 The use of dipole sources in the vicinity of a microresonator 
allows one to investigate the dependence 
of the optical modes on the size, shape and refractive index contrast associated with the resonator 
and its environment. 
Efforts to characterize the optical properties of WGMs in resonators 
 using modeling techniques include both excitation from a 
plane-wave 
beam \cite{Fujii:2005a,Fujii:2005b,Quan:2005a}, and excitation from an electric 
dipole source \cite{PhysRevA.13.396,chance1978molecular,Gersten:80,Gersten:81,Chew:1987a,PhysRevA.38.3410,Ruppin:82,Schmidt:12,pmid24921827}. 
The dipole sources can also serve as an effective analogue for 
fluorescent 
dyes or 
embedded nanoparticles that excite 
WGMs in microspheres \cite{Francois:13}. 

\subsection{Microspheres} 

An FDTD simulation of a polystyrene ($n = 1.59$) microsphere, 
with a diameter of $6$ $\mu$m, is carried out, as 
illustrated in Figure~\ref{fig:Sphere}. 
To generate an electromagnetic current to excite the WGMs, one or several 
electric dipole sources may be placed in the vicinity of the sphere.  
Here, an electric dipole source is used, which emits a Gaussian pulse 
with a central wavelength of $600$ nm, and a width of $5$ fs. 
In simulating the orientation of the dipole with respect to the surface of the sphere, 
both the tangential and radial cases are considered. 
Note that the pulse width is 
significantly narrower than the decay transition rate expected from a typical fluorescent source, such as Rhodamine 
dye, which is approximately 1--3 ns \cite{Barnes:92,pmid19950338}. 
A circular flux collection region with a diameter of $D_{\mathrm{flux}} = 2.58$ $\mu$m is 
placed a distance of $L_{\mathrm{flux}} = 240$ nm from the surface of the sphere, 
with its normal perpendicular to the surface.  
The power spectrum is then collected for wavelengths in the range $500$-$750$ nm. 
The spectrum is 
normalized to $P_0$, the dipole emission rate in 
an infinite bulk medium of the same refractive index as the surrounding medium. 

\begin{figure}[thp]
\begin{center}
\includegraphics[width=0.49\hsize]{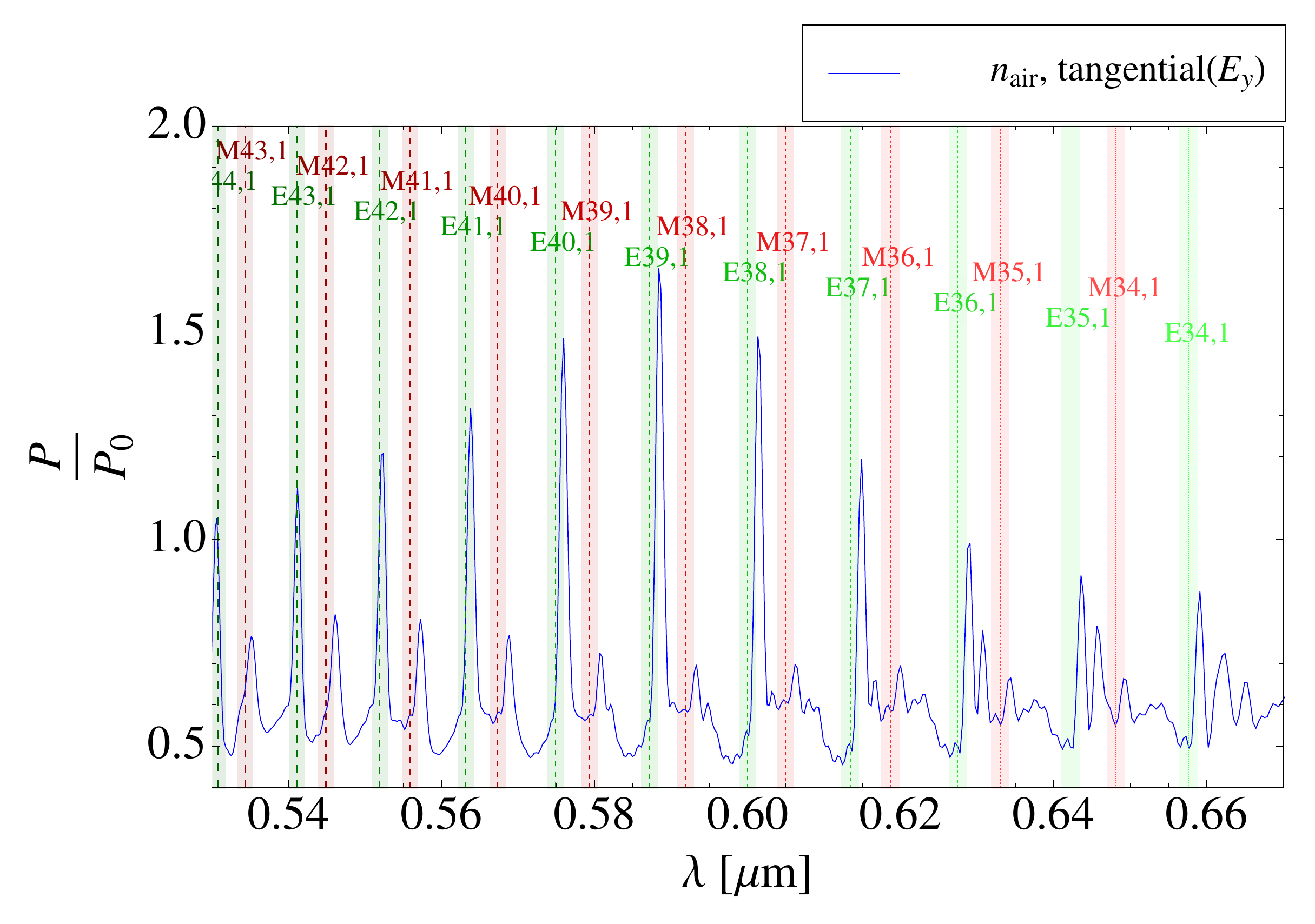}
\hspace{-3mm}
\includegraphics[width=0.49\hsize]{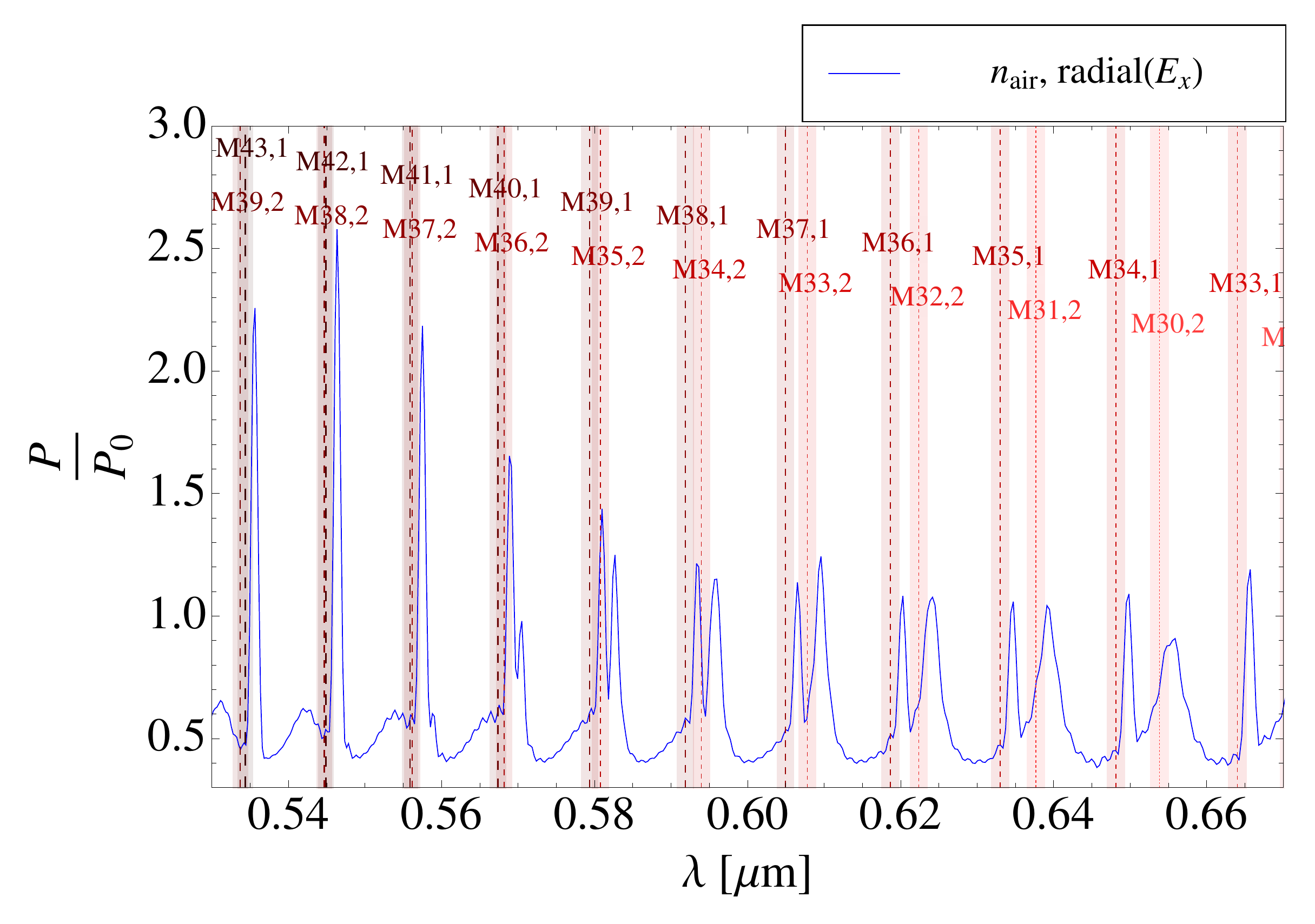}
\vspace{-3mm}
\includegraphics[width=0.49\hsize]{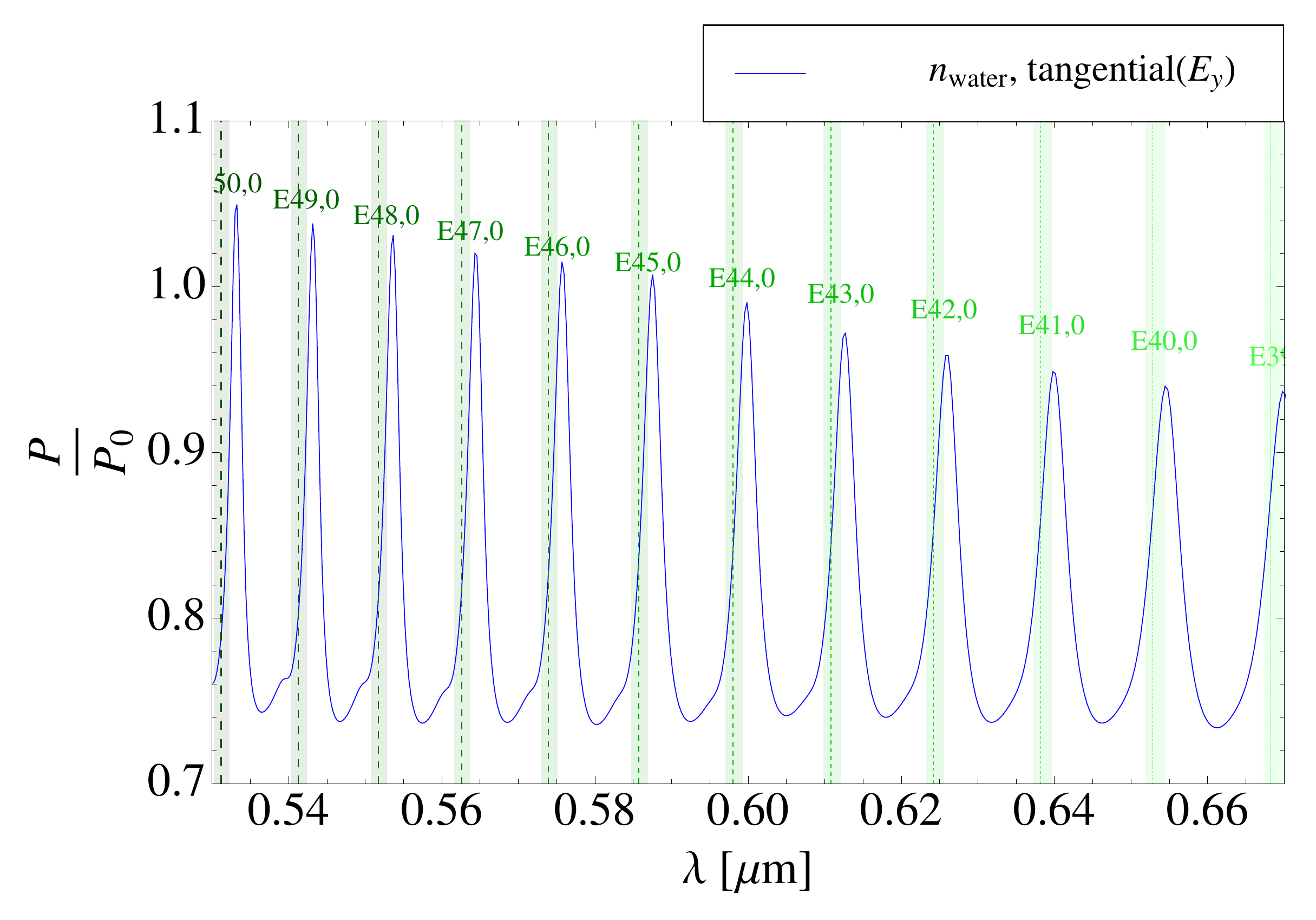}
\hspace{-3mm}
\includegraphics[width=0.49\hsize]{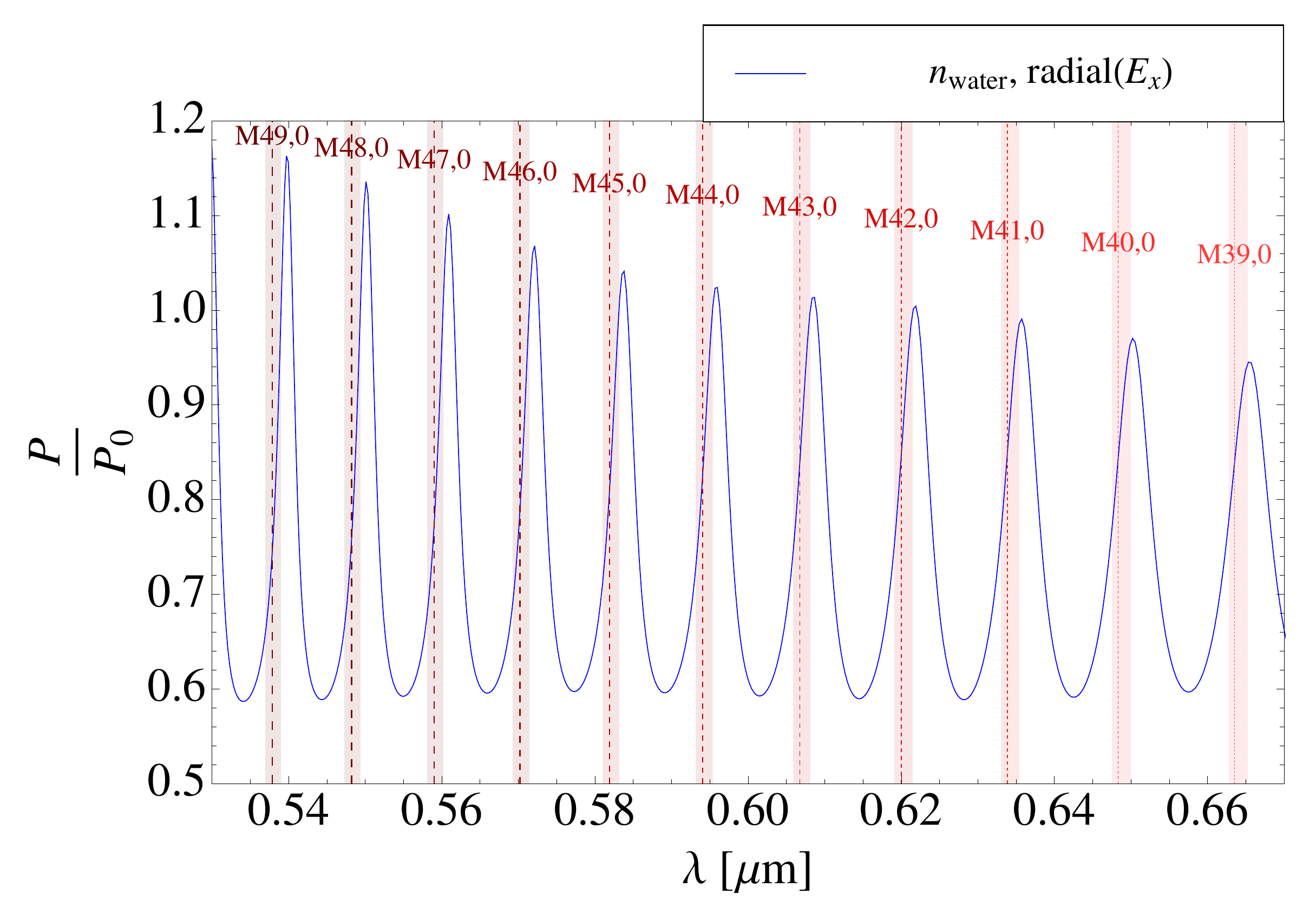}
\vspace{0mm}
\caption{\footnotesize{ FDTD simulation of the normalized power spectrum of a polystyrene microsphere, $6$ $\mu$m in diameter, with a surrounding medium of air. Whispering gallery modes are excited from a tangential (\textit{top left panel}) or radial (\textit{top right panel}) electric dipole source with a central wavelength of $0.6$ $\mu$m. 
The results for a surrounding medium of water are also shown for a 
tangential source (\textit{bottom left panel}) and a 
radial source (\textit{bottom right panel}). 
Vertical lines indicate predictions of the TE$_{m,n}$ (\textit{green}) and TM$_{m,n}$ (\textit{red}) modes derived from a typical analytic model \cite{Johnson:93,Teraoka:06a}, for azimuthal and radial mode numbers $m$ and $n$, respectively. The width of the bands indicates the systematic uncertainty in the positions due to the finite grid size of FDTD.}}
\label{fig:WGM}
%\end{center}
%\end{figure}
%
\vspace{1mm}
%\begin{figure}[t]
%\begin{center}
\includegraphics[width=0.5\hsize,angle=0]{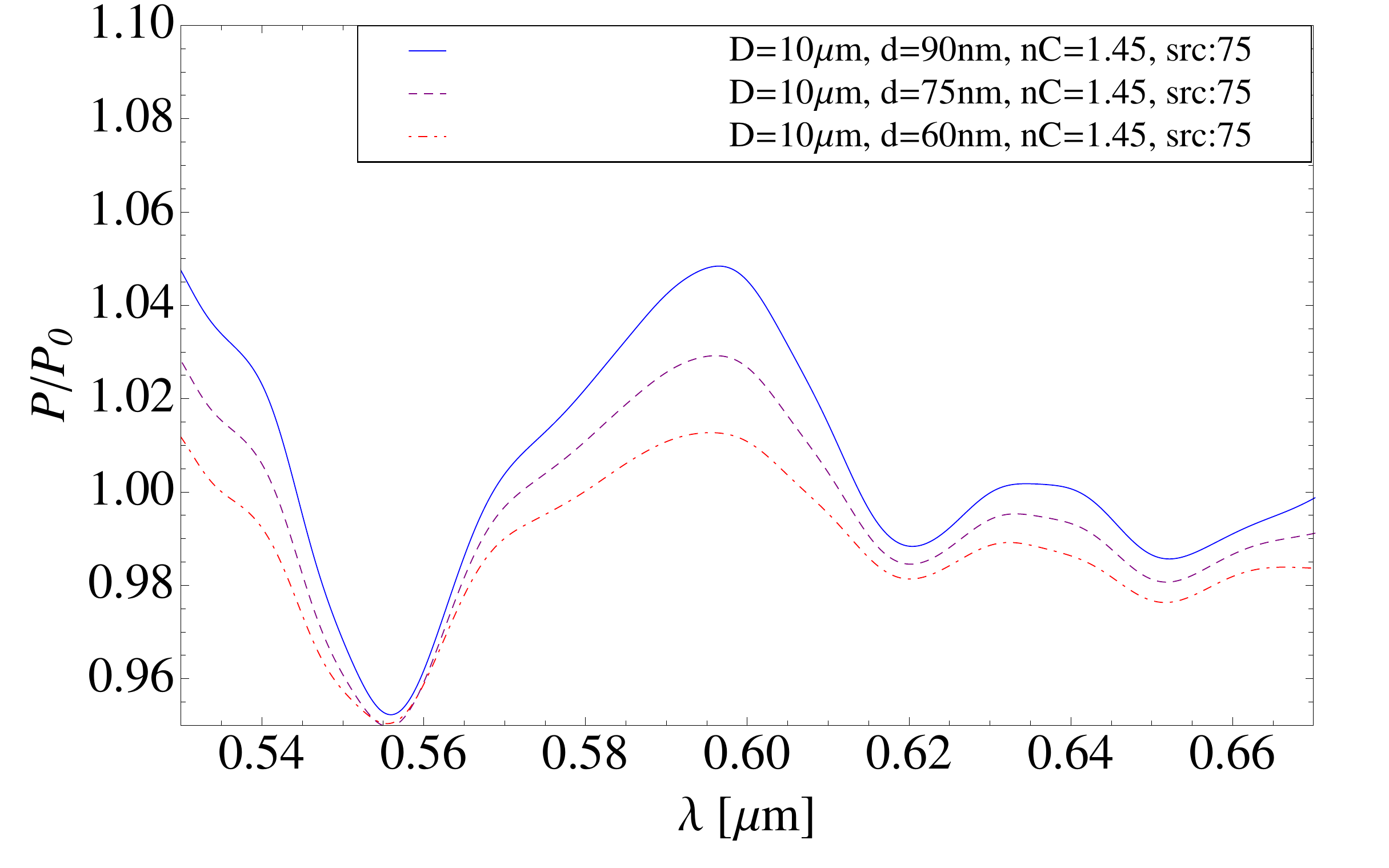}
\hspace{-3mm}
\vspace{-3mm} 
\includegraphics[width=0.49\hsize,angle=0]{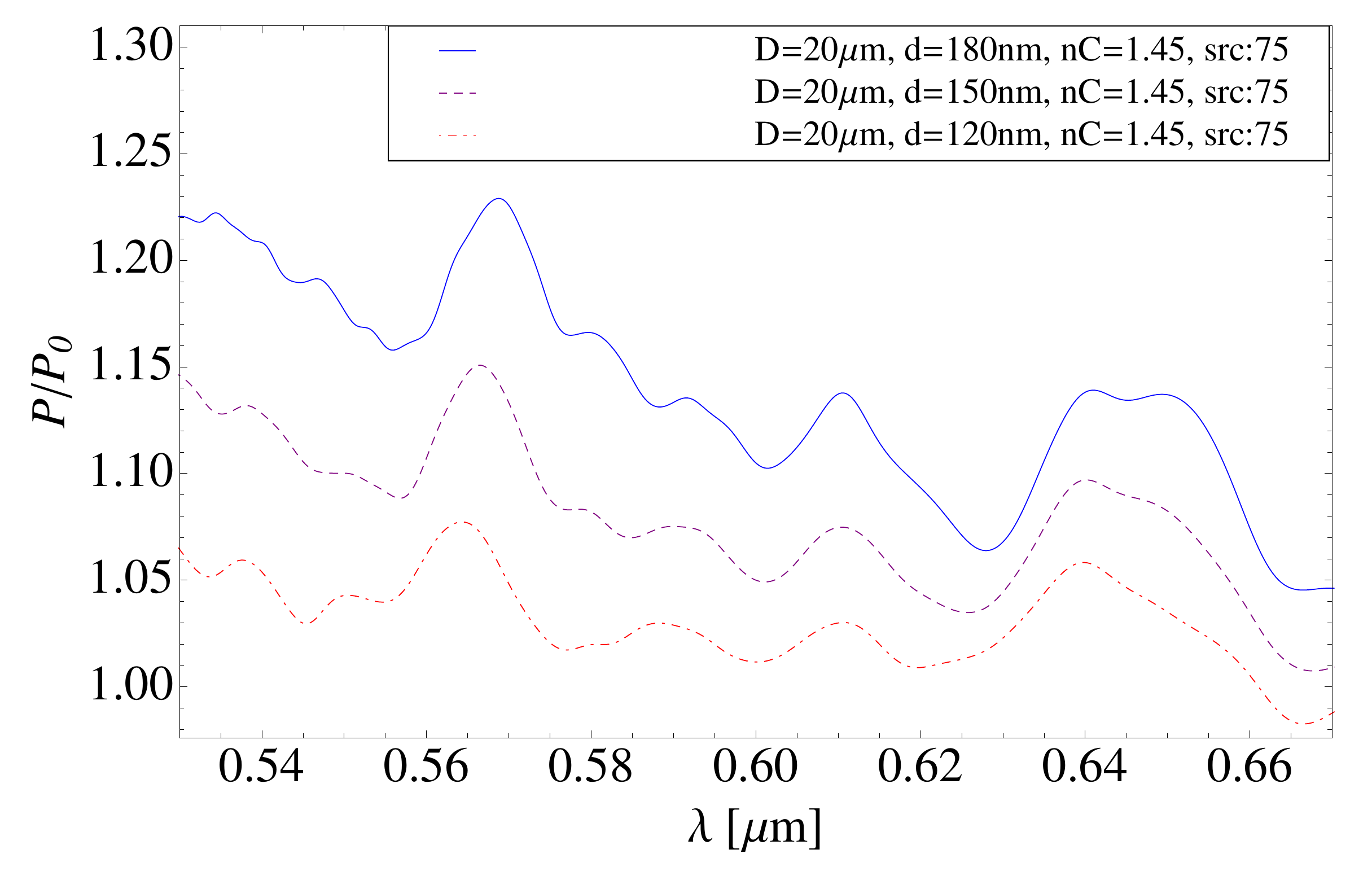}
\caption{\footnotesize{Power spectrum simulations of fluorescent microshells with a shell refractive index of $n=1.45$, and with a water medium ($n=1.33$) both inside and outside the shell. The modes of the shells are stimulated using an evenly-spaced distribution of $75$ electric dipole sources of random orientation, placed on the outer surface of the shell. \textit{Left panel}: the outer-shell diameter is $10$ $\mu$m with shell thicknesses of $60$, $75$ and $90$ nm. \textit{Right panel}: the outer-shell diameter is $20$ $\mu$m with shell thicknesses of $120$, $150$ and $180$ nm. A circular collection region (diameter: $2.58$ $\mu$m) is 
placed $240$ nm from the surface of the shell. }} 
\label{fig:shell}
\end{center}
\end{figure}

A comparison of the analytic 
model and the FDTD simulated spectrum for a microsphere 
is shown in Figure~\ref{fig:WGM}, for tangentially and radially 
oriented dipole sources in both air and water media. 
The WGM positions from the analytic model are shown 
as vertical bands, with TE$_{m,n}$ modes in green and TM$_{m,n}$ modes in red, 
for azimuthal and radial mode numbers $m$ and $n$, respectively. 
The width of each band indicates the estimate of the shift in the WGM positions due to the uncertainty in the 
sphere diameter, $6$~$\mu$m~$\pm\Delta x /2$. 
In this case, the spatial 
resolution is held fixed at the value $\Delta x=22$ nm. 
For the temporal resolution, 
the spectral density yields an uncertainty of $0.31$~nm. The two uncertainties are
 added in quadrature. 

\subsection{Microshells}

The FDTD method can be modified to consider a microshell resonator, where a hollow sphere is placed in a 
medium, and excited from a number of dipole sources. The shell may be of variable thickness, and in this 
investigation, 
is made from a higher refractive index material than the external medium. The interior of the shell is assigned 
the same refractive index as the surrounding medium. 
To approximate the effect of exciting the WGMs using a fluorescent coating, an evenly-spaced distribution  
of electric dipole sources is simulated on the outer surface of the microshell. The orientation of the dipoles 
is randomly generated, and the number of dipoles placed on the surface (in this case, $75$) 
is made large enough to provide an effective 
analogue of a medium of fluorescent emitters. 
Figure~\ref{fig:shell} shows the variation of the power spectrum due to changes in the shell thickness. 
In the left panel, an outer shell diameter of $10$ $\mu$m is used, with shell thicknesses of $60$, $75$ and $90$ nm 
considered. In the right panel, an outer shell diameter of $20$ $\mu$m is used, considering 
shell thicknesses of $120$, $150$ and $180$ nm. 
For a low refractive index contrast, a small sphere diameter and a small shell thickness, 
the WGMs are not easily resolvable. However, as the shell thickness increases, an improvement in  
 definition is observed in the spectrum structure. For spectra obtained from a larger shell diameter 
(shown in the right panel of 
Figure~\ref{fig:shell}) a peak structure begins to emerge as the shell thickness is increased. 
Note that the presence of the distribution of dipole sources serves to broaden the WGMs peaks by exciting 
a range of closely-spaced higher order modes.

\section{SUMMARY}

A customizable FDTD tool is established, and investigated in the context of whispering gallery mode generation 
in microresonators. 
The method can be tailored to more realistic scenarios, where the assumptions built into typical analytic models make direct comparison with experiment difficult. 
An important novel feature of FDTD is the ability to investigate the choice of source used for mode excitation, as-yet unexplored in the literature. Dipole sources may be placed at a variety of positions and orientations on the surface of the sphere, which
can serve as an analogue for nanoparticle coatings. 

The generation of whispering gallery modes in novel resonator structures represents an important application
 of the customizable FDTD approach. 
The FDTD tool is extended to explore configurations of microshell resonators, represented by a hollow sphere 
with a water medium placed both inside and outside, where the shell is made from a higher refractive index material. 
A distribution of randomly-oriented dipole sources is placed on the surface, analogous to a fluorescent dye 
coating. 

The FDTD tool presented here represents the first step in establishing a realistic optical configuration simulator, 
useful for facilitating a 
cost-effective approach to designing tailored optical resonators.  
Tuning the characteristics of the resonator for optimal power coupling and identifying the most prominent spectral features will help to reduce fabrication costs and aid the development of the next generation of biosensing tools.

\bibliography{pwest}  
\bibliographystyle{spiebib} 

\end{document}